\documentclass[]{aastex631}
\pdfoutput=1

\received{September 2, 2022}
\revised{November 8, 2022}
\shorttitle{Subneptune Runaway}
\shortauthors{Pierrehumbert}
\begin{document}

\title{The runaway greenhouse on subNeptune waterworlds}

\correspondingauthor{Raymond T. Pierrehumbert}
\email{raymond.pierrehumbert@physics.ox.ac.uk}

\author[0000-0002-5887-1197]{Raymond T. Pierrehumbert}
\affiliation{University of Oxford \\
Department of Physics \\
Oxford OX1 3PW, United Kingdom}


\begin{abstract}

The implications of the water vapor runaway greenhouse phenomenon for water-rich subNeptunes 
are developed. In particular, the nature of the post-runaway equilibration process for
planets that have an extremely high water inventory is addressed.  Crossing the
threshold from sub-runaway to super-runaway conditions leads to a transition from 
equilibrated states with cold deep liquid oceans and deep interior ice-X phases to
states with hot supercritical fluid interiors. There is a corresponding marked inflation
of radius for a given mass, similar to the runaway greenhouse radius inflation 
effect noted earlier for terrestrial planets, but in the present case the inflation
involves the entire interior of the planet. The calculation employs the AQUA equation
of state database to simplify the internal structure calculation.  Some speculations
concerning the effect of $\mathrm{H_2}$ admixture, silicate cores and hot vs. cold start
evolution trajectories are offered.  Observational implications are discussed, though
the search for the mass-radius signature of the phenomena considered is limited by
degeneracies and by lack of data.

\end{abstract}



\section{Introduction} \label{sec:intro}

Ever since the pioneering work of Ingersoll \citep{ingersoll1969runaway}, elaborated on in important ways
by \cite{kasting1988runaway} and \cite{nakajima1992study}, the runaway greenhouse phenomenon has played
a central role in thinking about planetary evolution. In particular, it is broadly taken as
defining the inner limit of the conventional liquid-water habitable zone.  Most attention
to date has focused on terrestrial planets, defined as planets with a rocky surface, but 
there is considerable current interest in the class of planets called subNeptunes, which may
well be the most common type of planet in our galaxy \citep{bean2021nature}. Hycean worlds
are subNeptunes with an $\mathrm{H_2}$ dominated atmosphere over a temperate liquid water ocean,
and have been proposed as a novel form of habitable world \citep{madhusudhan2021habitability}. 
Such planets are essentially a variant of the hydrogen-supported habitable states discussed by  
\cite{pierrehumbert2011hydrogen}, with more emphasis placed on application to subNeptune waterworlds.
The radiative results in \cite{pierrehumbert2011hydrogen}
have recently been confirmed by \cite{mol2022potential}, who also show that habitable conditions
can persist for a billion years or more.   Hycean worlds would runaway (or never
condense an ocean in the first place) if they are above the runaway threshold. 
Interest is further sparked by the detection of water in the hydrogen-dominated atmosphere of K2-18b
\citep{benneke2019water,tsiaras2019water}, which is estimated to be very near the threshold
instellation for runaway.
 
In this article we address the issue of how a water-dominated subNeptune equilibrates if
its instellation exceeds the runaway greenhouse threshold, and the discontinuous switch in
planetary characteristics that is expected for subNeptunes below vs. above the threshold.
The work reported in this article has close affinities with studies of subNeptune interior
structure \citep{mousis2020irradiated,aguichine2021mass, nixon2021deep} and of atmospheric
inflation in the post-runaway state on terrestrial planets by \cite{turbet2019runaway}.
Our work has the modest goal of bridging these two classes of study by showing how the interior
structure of a water-rich subNeptune depends on instellation, and on clarifying the way a
runaway greenhouse scenario plays out on a subNeptune. There has been a great deal of
previous work on the interior structure of subNeptunes (see \cite{nixon2021deep} for a
fairly comprehensive review of this body of work) and our results do not add particularly
to what is already known about the range of interior structures and their effects on the
mass-radius relation,apart perhaps from illustrating how the AQUA equation of state
database \citep{haldemann2020aqua} makes such calculations simple. However, little of that
work connects the interior structure to the planetary energy budget constraints imposed by
a self-consistent radiative-convective atmosphere; without such coupling  a connection cannot be made with the
runaway greenhouse phenomenon. The work of \cite{mousis2020irradiated} and
\cite{aguichine2021mass} on subNeptunes with supercritical water interiors is a notable
exception. The states treated there are essentially the same as the super-runaway states
treated in the present work, but the connection to the runaway greenhouse phenomenon is
only mentioned in passing in \cite{mousis2020irradiated}, and not explored in depth. The focus here is on clarifying the
dramatic transition in interior structure that occurs between planets with lower
instellation than the runaway threshold and planets with higher instellation, the nature
of equilibration of water-rich post-runaway subNeptunes, and the use of the runaway
threshold as a simply stated screening criterion for the existence of Hycean worlds. Our
results have some observational implications for the mass-radius distribution of the
population of weakly irradiated subNeptunes spanning the runaway greenhouse threshold.

There are two situations in which coexistence of gaseous and condensed phases restricts
the state of the system to the $p-T$ phase boundary between gaseous and condensed water.
First, within an atmosphere dominated by the gaseous phase, saturation can lead to the
formation of droplets or particles of the condensed phase, and so long as condensed phase
continues to form (e.g. through radiative cooling) the system will be restricted to the
phase boundary. The phase boundary is calculated by solving the Clausius-Clapeyron
relation for temperature as a function of the partial pressure of water $p_{H2O}$. This
will be called $T_{df}(p_{H2O})$. It is the usual meteorological definition of dew or frost
point. For the pure-steam atmospheres which are the central focus of this article,
$p_{H2O}$ is in fact the total pressure, and can simply be written as $p$. Alternately, if
the planet exhibits a layer of essentially pure condensed phase (an "ocean"), then the
vapor pressure of the gas phase in contact with the ocean surface is also restricted to
the phase boundary, though {\it within} the ocean, where there is no co-existing gas
phase, the $p-T$ trajectory is not constrained. In this case, the mass of the atmosphere
increases with the ocean surface temperature, as more of the liquid phase is converted to
atmosphere and the ocean becomes shallower.

The essence of the runaway greenhouse phenomenon is that, when the infrared radiating
level of the atmosphere lies in the portion where temperature is controlled by
$T_{df}(p)$, increasing the ocean surface temperature cannot increase the radiative
cooling to space, so the ocean surface continues to heat until some process intervenes to
allow the temperature of the radiating level to increase
(\cite{pierrehumbert2010principles}, Chapter 4). In the conventional picture the ocean
surface heats up until the entire ocean has evaporated into the atmosphere. At that point,
there is no more condensed reservoir to feed into atmospheric mass as the surface heats
up, and a noncondensing region governed by the noncondensing ("dry") adiabat forms
extending from the surface to the layer where condensation occurs. As the surface heats
up, the dry adiabatic region progressively eats into the bottom of the condensing region,
allowing the radiating region of the atmosphere to warm sufficiently to restore energy
balance \citep{boukrouche2021beyond}. (At very high temperatures, there is also some
contribution to the equilibration by the more widely recognized mechanism of short-wave
thermal radiation escaping through atmospheric window regions.) 

But what happens if essentially your entire planet is made of water, and it is impossible
to "run out" of ocean to evaporate? How, then, does the planet equilibrate in the
post-runaway stage? The key is that once the ocean surface temperature exceeds the
critical point, there is no longer a distinction between the fluid and gaseous phase, so
the atmospheric adiabat connects seamlessly to the supercritical water adiabat which
extends into the deep interior of the planet (and possibly connecting to high-pressure
fluid or ice phases there). Unlike the dew or frost point adiabat, which has no free
parameters once the composition is fixed, the supercritical adiabat (like the conventional
dry or noncondensing adiabat) has a free parameter, which can be regarded as its entropy,
or its temperature at a reference pressure level. To equilibrate, the adiabat needs to
heat up to the point where the radiation to space can increase. Equilibration then
proceeds as per the usual scenario, primarily by thinning the condensing layer, but also
with some contribution from the hot deep atmophere radiating through shortwave windows.
This picture will be made quantitative in the following.

\section{Methods} 
\label{sec:methods}

The central concept of how a runaway greenhouse plays out for a subNeptune will be
illustrated in the context of a highly idealized set of assumptions regarding the
planetary structure. Some ways in which deviations from this idealization might alter our
results will be discussed in Section \ref{sec:complications}. We assume the planet to be
completely composed of water, all the way to its center, and to be convective all the way
to the center, apart from a radiative stratosphere at the top of the atmosphere. These
assumptions constrain the $T(p)$ profile below the stratosphere to lie on the adiabat for
water substance in the appropriate phases. In the upper portions of the atmosphere, where
the non-condensing adiabat would be colder than the dew or frost point temperature
$T_{df}(p)$, the temperature is set to $T_{df}(p)$, which defines the phase boundary,

The non-condensing adiabat extending into the interior is computed using the AQUA equation of
state database for water, which provides tabulations of the adiabatic exponent as well as density
\citep{haldemann2020aqua}.  The equation of state covers all likely liquid or ice phases that may be
found in the interior. 

The depth of the condensing layer is determined by the radiation balance of the planet.
Although the analysis in \citep{boukrouche2021beyond} was done for a terrestrial planet with
a rocky surface, the dominant radiation escaping to space comes from the atmosphere and
not from the surface, and so the results can be applied to subNeptunes. The atmospheres 
considered were shallow compared to a typical planetary radius, so the "surface gravity"
in this study is essentially the same as the gravity at the level of atmosphere from which
radiation escapes to space, which is the relevant "surface gravity" to use in the context
of a subNeptune, which has no distinct surface anywhere near the radiating layer. 
For surface gravity 10$m/s^2$, the limiting outgoing longwave radiation ($OLR$) for a
saturated pure steam atmosphere, which defines the runaway greenhouse threshold, is
approximately 280 $W/m^2$ (as read off the flat portion of Fig. 8 in  \citep{boukrouche2021beyond}).
For absorbed stellar radiation ($ASR$) above this value, the deep atmosphere will continue
to warm until the noncondensing adiabat extends sufficiently far into the condensing region
to allow the radiating level to warm sufficiently that the $OLR$ increases so as to balance the
$ASR$.

The corresponding grey-equivalent radiating temperature to the runaway threshold flux is
265K; based on the dew-point formula. This corresponds to a radiating pressure of 333Pa.
However, because of radiation through fairly transparent window regions a considerable
amount of flux originates somewhat deeper in the atmosphere than this layer. Based on the
surface temperature at which the OLR begins to rise in Fig. 8 of
\citep{boukrouche2021beyond}, and the corresponding profiles shown in Fig. 3 of that work,
the actual base of the condensing layer in a slightly super-runaway state (i.e. one for
which the $ASR$ slightly exceeds the 280$\mathrm{W/m^2}$ runaway threshold) is closer to
1000 Pa, and this is the value we shall take as our estimate of the base of the condensing
layer for the slightly sub-runaway and slightly super-runaway cases. The corresponding
temperature on the dew point adiabat is 280K, and thus slightly above freezing. This point
is a key parameter since it determines the pressure at which the $p-T$ trajectory
intersects the phase boundary, and therefore provides the starting point for continuing
the adiabat into the planetary interior. This prescription allows a simplified treatment
of the atmospheric radiative transfer, avoiding the complexity of the more precise use of
a full radiative-convective calculation such as employed in \cite{mousis2020irradiated}.
The limiting $OLR$, upon which our analysis is based, is only weakly dependent on gravity
\citep{pierrehumbert2010principles}.

$T_{df}$ becomes arbitarily cold as $p\rightarrow 0$, but the temperature decrease is
halted by the formation of a radiative-equilibrium stratosphere above some level, which is
warmer than condensation threshold. Stratospheres are maintained both by absorption of
upwelling thermal radiation from below, and by {\it in situ} absorption of incoming
stellar flux. As is common practice, we will represent the stratosphere as an isothermal
layer, though optically thick stratospheres are not generally isothermal. For illustrative
purposes, it will be assumed that the stratosphere is thin enough that it does not
significantly affect the top-of-atmosphere radiation balance. This can be a poor
approximation for very highly irradiated planets, but our focus here is on planets that
are near or below the runaway threshold. As the $ASR$ increases, the stratospheric
temperature will increase both because the $OLR$ needed to balance the $ASR$ increases
(implying an increase in the infrared flux heating the stratosphere from below) and
because the {\it in situ} heating by stellar absorption increases.

\begin{figure}[ht!]
\plotone{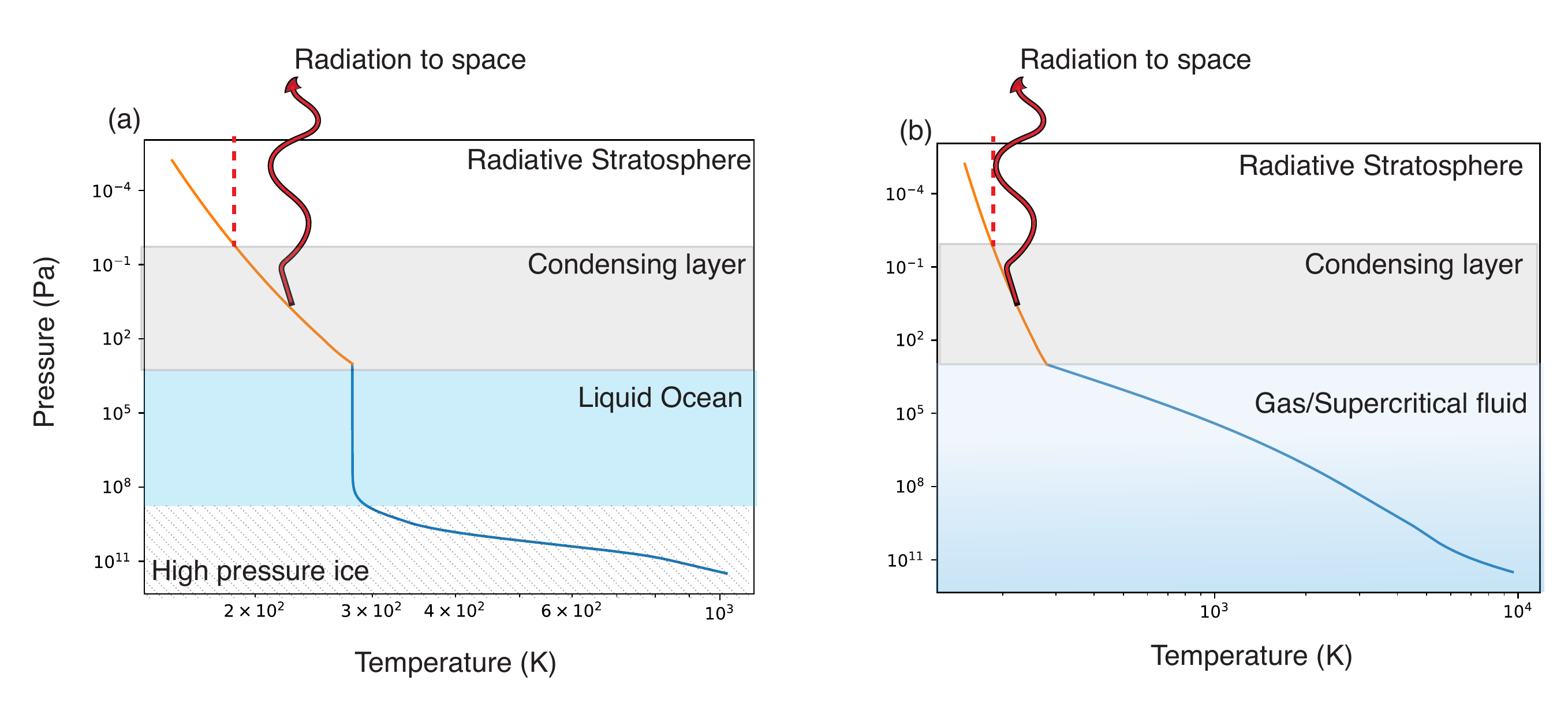}
\caption{Planetary structure for a slightly sub-runaway (a) and slightly super-runaway (b) absorbed stellar radiation. \label{fig:profiles}}
\end{figure}

When the $ASR$ is just below the runaway threshold, a liquid ocean can be
maintained, and the bottom of the condensing layer is at the liquid ocean interface. In this
case the saturated condensing layer is in contact with the surface, and given our estimate of the
base of the condensing layer only a thin water vapor atmosphere would remain, with surface pressure
1000Pa (or less at lower $ASR$ and lower surface temperature).   At greater depths,
the adiabat continues into the liquid adiabat, which is not constrained to be on a phase boundary.  This
situation is depicted in Fig.~\ref{fig:profiles}a, for a case where $ASR$ is just slightly below the runaway
threshold.  When the $ASR$ is above the threshold, a liquid ocean cannot be maintained, and instead the interior of the planet must heat up until water becomes supercritical at some point. In this case, the bottom of the
condensing layer is at the boundary of a noncondensing (dry adiabatic) region, as in \cite{boukrouche2021beyond}.
This adiabat is subcritical at the point of contact but continues seamlessly into the supercritical fluid
adiabat.  The situation is depicted for $ASR$ slightly above the runaway threshold in  Fig.~\ref{fig:profiles}b.

The situation described above corresponds most closely to a hot-start case, in which the planet initially
forms with a hot interior, which drives a deep convective layer as the heat escapes.  The calculation
then describes the state the planet attains as the interior cools down sufficiently that the the top of
atmosphere comes into balance with $ASR$, though while there is still enough interior heat flux to
maintain convection. Super-runaway planets will equilibrate in a supercritical fluid state as they cool,
whereas sub-runaway planets will eventually condense a liquid water ocean in the interior. The hot
start scenario also applies to the period of declining $ASR$ characteristic of the pre-main sequence
stage of low mass M and K stars.  

The cold-start case, in which the planet starts in a sub-runaway condition with a 
cool interior and liquid ocean, but then crosses the runaway threshold owing to luminosity increase
of its star, is considerably more complex owing to the difficulty of mixing heat downward against a stable
buoyancy gradient.  In such cases, the assumption of a fully convective interior is clearly inappropriate. 

Once $T(p)$ has been determined, the mass-radius relation can be computed in the usual
fashion. The equation of state yields the density $\rho(p)$. The hydrostatic relation
yields $dp/dr$ given $\rho$ and the local acceleration of gravity $g(r)$. $g(r)$ is in turn given by
Newton's law of gravitation, in term of the mass $M(r)$ contained within a shell of radius
$r$, while $dM/dr$ is determined by the density $\rho(p)$. This yields a pair of coupled
ordinary differential equations which can be integrated from $r=0$ given a guess at $p_0$,
the pressure at the center of the planet. The equations are integrated outward until a
specified low pressure $p_\mathrm{top}$ is reached. This procedure yields a planetary mass
$M(p_0)$ and radius $r(p_0)$, so that the mass-radius relation can be mapped out
parametrically by varying $p_0$. The value of $p_\mathrm{top}$ should in principle be
determined by the pressure at which the limb of the atmosphere becomes opaque to the
wavelength band in which the planet is being observed, but to specify it in this way would
become dependent on the nature of the observation. In practice, the computed radius is
relatively insensitive to the precise value chosen for $p_\mathrm{top}$, because pressure
decays approximately exponentially in the outer atmosphere, whence the estimated radius
depends only logarithmically on $p_\mathrm{top}$. The resulting radius can be taken as a
baseline from which apparent radius as a function of wavelength can be computed in the
analysis of transit depth observations. In results presented here, we use $p_\mathrm{top}
= 1000 Pa$. All references to "surface gravity" below refer to the gravitational
acceleration computed at this pressure, which is close to the gravitational acceleration
at the infrared radiating level.

\section{Results}

The curves shown below the stratosphere in Fig.~\ref{fig:profiles} are the actual adiabats computed from
the AQUA equation of state, with corresponding phases indicated. 

The pressure that corresponds to the center of the planet depends on the planet's mass,
and will be discussed shortly. For slightly sub-runaway conditions, there is a deep, cool
liquid ocean. This is nearly isothermal, even if fully convective, because the adiabatic
exponent for liquid water is very small. At higher pressures, there is a transition to
Ice-X, which has a higher exponent, leading to a layer of increasing temperatures. For
lower $ASR$ (not shown), the temperature of the condensed surface goes down, and the
surface pressure (marking the lower boundary of the condensing atmosphere) goes down
correspondingly. It takes very little reduction in $ASR$ for the ocean surface to freeze,
leading to an increase in albedo and further reduction in surface temperature and
pressure. This ice-albedo feedback is especially important for the case of G stars, but less so for
the case of M or K stars whose infrared-rich spectrum leads to lower ice albedo
\citep{joshi2012suppression,shields2013effect,von2013dependence}. A computation based
strictly on the AQUA adiabat would lead to an Ice-I shell transitioning to higher pressure
ice phases in the deeper interior, but in reality, the rigidity and low thermal
conductivity of the shell would inhibit convection, so even a slight interior heat flux
would lead to a liquid water ocean under a relatively thin ice shell, as is well known
from Snowball Earth calculations (e.g. \cite{li2011sea}). This is also the state of
Europa, so the resulting subNeptune could be considered a super-Europa.

From a hot start, the sub-runaway equilibrium situation depicted in
Fig.~\ref{fig:profiles}a would take some time to emerge, because the interior would only
gradually cool as the excess heat of formation in the supercritical water deep interior
radiates away to space. Liquid or ice layers would only start to accumulate once the
interior cooled sufficiently for condensate to persist in the deep interior.
At intermediate times, the liquid and Ice-X deep interior in the equilibrium state would
be replaced by hot supercritical water, leading to a corresponding increase in the radius
of the planet relative to the equilibrium state. Modeling the thermal evolution of
subNeptunes is beyond the scope of the present work.

When the $ASR$ is slightly above the runaway threshold, the gas and supercritical adiabat (Fig.~\ref{fig:profiles}b)
have a uniformly large adiabatic exponent, leading to a very hot ($\approx 10000K$ ) interior, which will reduce
the density and inflate the planet. The contrast between Fig.~\ref{fig:profiles}a and Fig.~\ref{fig:profiles}b 
indicates the dramatic transition that occurs for subNeptunes below vs. above the threshold. In the super-runaway
case, condensation and consequent formation of water clouds occurs only in a thin layer near the top of the atmosphere,
even if the $ASR$ only slightly exceeds the runaway threshold.  This results from the very nature of the post-runaway
equilibration, and applies for terrestrial planets as well. 

At higher $ASR$ (not shown) the gas/supercritical adiabat intersects the condensing adiabat at lower pressures and
somewhat lower temperatures, so as to allow the greater radiation from the hot noncondensing layer to balance
$ASR$.  Since the reduction of temperature with pressure on $T_{df}$ is slight, this leads to a hotter adiabat and yet hotter
supercritical interior temperatures.  For sufficiently large $ASR$ the increase in stratospheric temperature
and deepening of the noncondensing layer would eliminate the condensing layer altogether. This is a familiar
situation for highly irradiated subNeptunes such as GJ1214b, for which water never condenses \citep{miller2010nature}.
As a simple estimate, if one estimates the stratospheric temperature by grey-gas skin temperature, GJ1214b would
have an upper-atmosphere temperature of 506$K$, well above the temperature of the condensing layer in a post-runaway
steam atmosphere.

The radius, surface gravity and central pressures are given for various planetary masses
in Table~\ref{table:massrad}, for the slightly sub-runaway and slightly super-runaway
cases. As expected, the super-runaway cases are very inflated relative to the sub-runaway
cases, though the proportional inflation decreases with increasing mass, as the high
pressure interior in the hot case becomes less compressible. The inflated runaway cases
are similar to the runaway greenhouse radius inflation discussed by
\cite{turbet2019runaway}, except that they involve the entire interior of the planet, and
are the same kinds of states as the supercritical water states discussed comprehensively
in \cite{mousis2020irradiated}. The 300K equilibrium temperature case in Fig. 2 of
\cite{mousis2020irradiated} is closest to being comparable to our marginally super-runaway
states, and our computed super-runaway radii are reasonably consistent with the results
there. For example, for $3M_\oplus$ our radius is 2.47$r_\oplus$ vs. 2.7$r_\oplus$ in
\cite{mousis2020irradiated}; for $9M_\oplus$ it is 2.9$r_\oplus$ vs. 3.4$r_\oplus$. It is
to be expected that the cases in \cite{mousis2020irradiated} are somewhat inflated
relative to our marginally super-runaway states, since 300K equilibrium temperature
corresponds to an implied ASR of 460 $\mathrm{W/m^2}$, vs. the 280 $\mathrm{W/m^2}$
assumed in our marginally super-runaway states.

Table~\ref{table:massrad} also shows the surface gravity for each case. The runaway
threshold is weakly dependent on surface gravity \citep{pierrehumbert2010principles},
because the hydrostatic relation implies that for low gravity it takes less pressure at
the bottom of the condensing layer to make it optically thick, though the dependence is
reduced by the pressure-dependence of opacity. Low gravity reduces the $ASR$ threshold for
runaway and high gravity increases it. The super-runaway cases below 9$M_\oplus$ all have
lower gravity than the nominal 10$m/s^2$ we assumed in the calculation, so in fact they
exceed the runaway threshold by somewhat more than assumed in the calculation;
accordingly, the interiors would be somewhat warmer than computed and the planet would be
somewhat puffier. A fully self consistent calculation would need to iterate on surface
gravity and $ASR$ threshold, but except for very low masses the effect is not likely to be
great. It is interesting that the runaway actually helps to reinforce itself, through
reduction in surface gravity. The sub-runaway cases can also have reduced gravity for the
lowest mass cases, but by 9$M_\oplus$ the gravity is 40\% above the nominal value assumed,
which would render the planet somewhat more sub-runaway than assumed, yielding a somewhat
cooler interior. Similarly to the super-runaway case, the gravity feedback reinforces the
state, pushing the planet further into the sub-runaway regime.

\begin{table}
\begin{tabular}{c|ccc|ccc}
\toprule
 &\multicolumn{3}{c|}{Sub-Runaway}&\multicolumn{3}{c}{Super-Runaway}\\
$M$ & $r$ &$g$ & $p_0$ &$r$ &$g$& $p_0$ \\
\toprule
1.5 & 1.57 & 5.9& 86  & 2.29 &2.8 & 45  \\
3.0 & 1.90 & 8.3 & 170 & 2.47 & 4.9 & 103 \\
6.0 & 2.26 & 11.5 & 337 &2.74 & 7.9 & 234 \\
9.0 & 2.5  & 14.1 & 522 & 2.9 & 10.3 & 380 \\
\end{tabular}
\caption{Radii, surface gravity and pressure at planet's center for planets of various masses,
for planets slightly below and slightly above the runaway greenhouse threshold. $M$ in Earth masses,
$r$ in Earth radii, $g$ in $m/s^2$, and $p_0$ in GPa.}
\label{table:massrad}
\end{table}

\begin{figure}[ht!]
\plotone{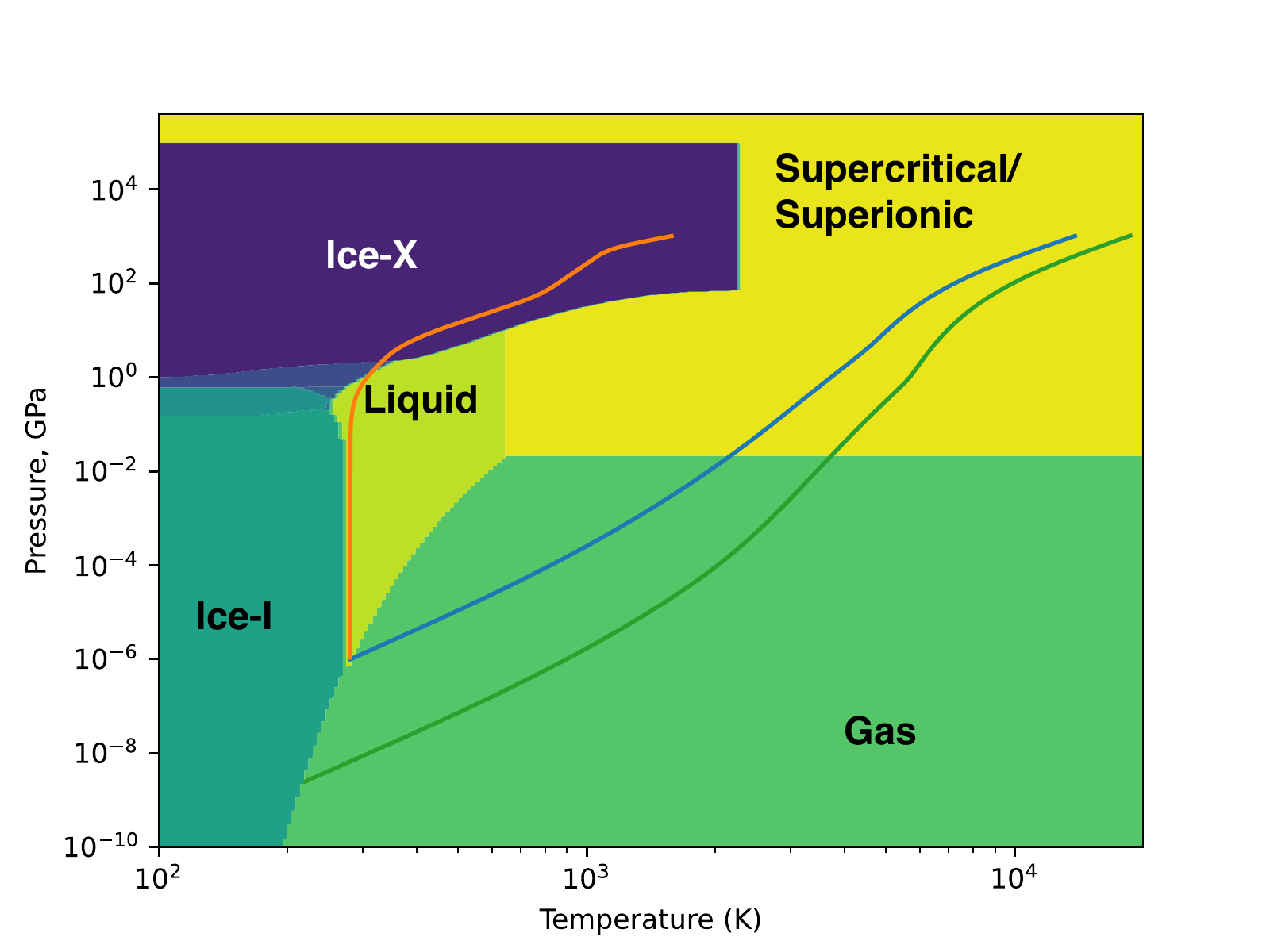}
\caption{Sub-runaway and super-runaway adiabats plotted on the AQUA phase diagram. Note that in this plot
the pressure axis is increasing towards the top of the figure.  At pressures below the point where
the adiabats intersect the gas/liquid or gas/ice-I phase boundary, the thermodynamic trajectory follows
the phase boundary until the stratosphere (barely present in this plot) is encountered.   \label{fig:PhaseDiagram}}
\end{figure}

In Fig.~\ref{fig:PhaseDiagram} we plot the adiabats on the AQUA phase diagram, including
a case of a hotter adiabat corresponding to $ASR$ somewhat in excess of the runaway
threshold. Note that the boundary between the gas and supercritical regions is not
actually a phase transition, but only indicates a switch in the equation of state employed
in the AQUA compilation. AQUA does not distinguish between supercritical and super-ionic
states, but an examination of the phase diagram in \cite{millot2019nanosecond} shows that
the super-runaway adiabats are too hot and too low pressure to venture into the
super-ionic ice region. This contrasts with the higher pressure adiabats starting from
colder upper atmosphere conditions, shown for Uranus and Neptune in
\cite{millot2019nanosecond}. The adiabats shown in the figure are computed out to
1000$GPa$, but the pressures given in Table~\ref{table:massrad} determine where the
adiabats actually terminate for planets of a given mass. In all cases, the super-runaway
planets terminate in a supercritical fluid region.

\section{Complications, caveats and extensions}
\label{sec:complications}

Our most vulnerable assumption is that the entire atmosphere and interior below the
condensing region is convective. The low thermal and radiative diffusivity of the deep
interior may mean that it takes very little interior heat flux to maintain convection
there, but stellar flux attenuates quite strongly with depth in a steam atmosphere,
particular in the case of M stars (see e.g. Fig. 5.14 in
\cite{pierrehumbert2010principles}). Analytic grey gas solutions
\citep{guillot2010radiative,pierrehumbert2010principles} show that in the absence of
interior heat flux the atmosphere asymptotes to a deep hot isothermal region below the
level to which appreciable stellar flux penetrates. The temperature of the radiative
layer increases like the fourth root of the infrared optical depth measured at the level
to which stellar flux penetrates (\cite{pierrehumbert2010principles},Eq.~4.56). Insofar as atmospheres (including water vapor) are
much more optically thick in the thermal infrared than at shorter wavelengths, the radiative
layer is generally very hot, though less so for M stars than for hotter stars. 
Assuming convection resumes in the deep interior, the isothermal radiative layer would set the interior
adiabat to a lower (though generally still supercritical) temperature than the adiabat connecting without
interruption to the bottom of the condensing layer.  Such states would reduce the inflation of runaway
subNeptune waterworlds, with the magnitude of the effect increasing as the depth of the radiative layer increases.   

The same solutions show that convection can be restored by sufficient interior heat flux if the absorption
coefficient of the atmosphere increases sufficiently rapidly with pressure. Nearly isothermal 
radiative layers are familiar from studies of the atmospheric structure of GJ1214b \citep{miller2010nature},
though the hot upper atmosphere of such highly irradiated planets makes it easier to suppress convection
and planets closer to the runaway threshold likely have thinner radiative layers, if any. We do not
attempt to estimate the amount of interior heat flux needed to yield a fully convective atmosphere because
the threshold is highly sensitive to the opacity behavior of water vapor at high temperatures and pressures.
For the most part, these spectroscopic properties -- particularly the crucial continua -- have been validated
only for relatively Earthlike conditions. 

According to current observations K2-18b has an $\mathrm{H_2}$ dominated outer atmosphere,
with possibly water as a minor constituent. Given the many ways planets can accrete both
$\mathrm{H_2}$ and water, this may be a common situation among subNeptunes. Adding a
noncondensible background gas which has fairly strong infrared opacity generally brings
down the $OLR$ for a given ocean surface temperature, but does not change the threshold
for a runaway, which is largely governed by the pure-steam limit. For the case of addition
of noncondensible $\mathrm{CO_2}$, this can be seen in the $OLR$ curves computed in
\cite{wordsworth2013water}. Because of its strong collisional opacity, $\mathrm{H_2}$ has
a similar effect, as was shown by \cite{koll2019hot} for the case of a water saturated
$\mathrm{H_2-H_2O}$ atmosphere assumed to be on the moist adiabat, though in that case
there are thermodynamic as well as opacity effects of the $\mathrm{H_2}$ background. Thus,
in a sub-runaway case, the addition of a noncondensible greenhouse gas to the atmosphere
of an ocean world increases surface temperature and moisture content of the lower
atmosphere, but does not trigger a runaway. In their analysis of habitability of Hycean
worlds, \cite{madhusudhan2021habitability}, incorporated the effects of
$\mathrm{H_2}$ and $\mathrm{H_2O}$ opacity in the determination of ocean surface temperature, but because the
$\mathrm{H_2O}$ concentration was assumed fixed (rather than increasing with temperature
as it would in equilibrium with an oceanic reservoir), the calculation did not factor in
the effect of the runaway greenhouse threshold, in contrast to \cite{koll2019hot}, who
allowed for the $\mathrm{H_2O}$ source arising from equilibration with an oceanic
reservoir. In any event, the increased ocean temperature due to additional opacity sources
would lead to a somewhat larger planet relative to the equilibrated all-water sub-runaway
solutions we have exhibited.

The preceding remarks imply that for a super-runaway case, a planet with a moderately
thick $\mathrm{H_2}$ envelope over a massive water ocean would run away until the
atmosphere became steam dominated. In a hot start condition with large water inventory and
moderate $\mathrm{H_2}$ inventory, the planet would equilibrate in the supercritical
post-runaway state we have exhibited, with the $\mathrm{H_2}$ diluted by the water
inventory. "Moderate" in this context means that the planet's $\mathrm{H_2}$ inventory is
small compared to its $\mathrm{H_2O}$ inventory. The reason $\mathrm{H_2}$ would become
diluted into the supercritical water interior is that the dissolution of gases in
supercritical fluids is no longer constrained by Henry's law solubility; rather, the
gases become completely miscible with the supercritical fluid \citep{budisa2014supercritical}.
We have been unable to find published experiments specifically on the case of $\mathrm{H_2}$
mixing with supercritical water, but experiments with $\mathrm{CH_4}$ solubility in water show that
gases can become nearly miscible with high-temperature water even before the critical point is 
reached \citep{pruteanu2017immiscible}.

It is of interest to explore the consequences of our results for the state of K2-18b. The
fact that K2-18b does not have a steam-dominated outer atmosphere means that either cloud
effects have put it into a sub-runaway state with a deep liquid ocean, or (if it has
super-runaway $ASR$) has a small water inventory relative to its $\mathrm{H_2}$ inventory.
The 9$M_\oplus$ case in Table~\ref{table:massrad} is close to the estimated mass of
K2-18b. The radius of an all-water sub-runaway state is 2.5$r_\oplus$, which only slightly
exceeds the observed 2.37$r_\oplus$ radius, but the upper bound on the radius estimate is
2.59$r_\oplus$, bracketing our sub-runaway state. If the planet can be kept sufficiently
cold through high albedo, the addition of a modest $\mathrm{H_2}$ envelope could dilute
the thin water vapor atmosphere and cold-trap water at low levels, yielding the observed
$\mathrm{H_2}$ dominated atmosphere. A super-runaway waterworld state, at 2.9$r_\oplus$ is
well in excess of the observed radius. A silicate core could perhaps bring the radius down
enough to to match observations, but the super-runaway waterworld state is already ruled
out by the observed atmospheric composition, based on our inference that a water-dominated
super-runaway state would have a water-dominated outer atmosphere. This line of argument
lends further weight to existing observational work tending to rule out the existence of a
liquid ocean. Specifically, \cite{scheucher2020consistently} concluded that for K2-18b
equiibration with an interior pure water reservoir would yield a high molecular weight
outer atmosphere and hence a transit depth spectrum incompatible with the observations.
Similarly, \cite{blain20211d} and \cite{charnay2021formation} concluded that transit-depth
observations are most consistent with a low metallicity $\mathrm{H_2}$ dominated
atmosphere with water present as a minor constituent. The speculative possibility remains
that an as-yet unidentified mechanism could yield high albedo clouds that reflect enough
instellation to put the planet in a sub-runaway state and permit a liquid ocean and
condensed interior, but such a cloud deck would need to be deep enough in the atmosphere
so as to not suppress the amplitude of the transit depth spectrum.

SubNeptunes may have a silicate/iron core.  In principle it would be easy to generalize our calculation by terminating
the AQUA adiabats at the core interface, using core radius and mass based on established compressibility estimates.  For
the hot super-runaway cases, though, the calculation is greatly complicated by the fact that the temperature at the
silicate boundary would be high enough to create a magma ocean and associated silicate vaporization, unless 
the supercritical water envelope is very thin (less than about 10$MPa$).  Adequate treatment of the problem in
that case would require consideration of the effects of supercritical water exchange with the silicate melt, and
the stabilization of the atmosphere against convection by the high molecular weight of silicate vapor relative to
water, such as considered for $\mathrm{H_2}$ envelopes by \cite{misener2022importance,dorn2021hidden}.  Incorporation of a moderate
silicate/iron core would contract the planet somewhat, but not alter our fundamental conclusion that for water-rich
subNeptunes crossing the runaway threshold effects a profound change in the interior state of the planet.

A subNeptune waterworld could undergo a cold-start runaway if it cooled off sufficiently,
early in its life, to form a condensed interior, but then the host star luminosity
increased sufficiently to cross the runaway threshold. This scenario is most likely for G
or higher mass main sequence stars, which have a short pre main-sequence stage but become
significantly more luminous on the main sequence on a time scale of a few billion years or
less. For a cold start runaway, the atmosphere would still heat up to the depth to which
significant stellar flux penetrates. For the cold start, there would be little heat flux
escaping the interior, and hence nothing to drive convection in the interior. This would
result in a hot (and possibly supercritical) isothermal upper layer in contact with a cold
liquid or ice boundary. As the isothermal layer radiated into the ocean or ice, it would
progressively advance toward the center of the planet through evaporating part of the
condensed region and heating it to match the isothermal layer temperature. The energy
required to do this would cool the lower part of the radiative layer, and the energy
needed to support further advance of the front must be supplied by radiative or molecular
diffusion through the deep isothermal radiative layer. The lack of opacity data for hot
high pressure water poses a severe challenge for quantifying the rate of advance of the
isothermal layer towards the planet's center, but the time scale for equilibration of the
planet's full interior is certain to be longer than in cases where heat transport is
dominated by convection. It is an open question how far the isothermal layer can progress
during the ages of observed subNeptunes, and this topic requires further study, as does
the general issue of thermal evolution of subNeptune waterworlds.

\section{Discussion}
\label{sec:discussion}

We have argued that for water-dominated subNeptunes, crossing the runaway greenhouse instellation threshold
leads to a profound transition in the interior state of the planet, from one dominated by liquid water and 
high pressure ice at moderate temperature to one dominated by supercritical water.  Although our calculations
have been performed explicitly only for the case of pure water composition, we have argued that the fundamental
conclusion would not be greatly altered by a moderate silicate/iron core or the presence of a minor proportion
of $\mathrm{H_2}$.

The transition is manifest in an inflation of the planet of a given mass when the
threshold is exceeded, similarly to the runaway greenhouse radius inflation discussed for
more limited ranges of depth by \cite{turbet2019runaway}. For a universe of pure water
subNeptunes, one would expect a discontinuous population of planets on the mass-radius
diagram for planets below vs. above the runaway threshold. Any attempt to confirm this
prediction against data will run up against the usual degeneracies in interpretation of
the mass-radius data: those associated with possible presence of a silicate/iron core and
those associated with presence of a hydrogen envelope. To some extent the hydrogen
degeneracy can be ameliorated through observations of atmospheric composition, but this
only constrains the outer portions of the atmosphere. We have argued that if a planet has
a composition in which hydrogen is a minor constituent relative to water, then in a
super-runaway state the hydrogen would be diluted into the massive supercritical water
layer, leaving a steam-dominated atmosphere that is incompatible with the observed
hydrogen-dominated atmospheres of planets such as K2-18b. Further, our results position
the runaway threshold of absorbed stellar radiation as a simple screening criterion for
the existence of Hycean worlds, since any candidate planets that exceed the runaway
threshold would evolve into a state with a supercritical water interior (as in
\cite{mousis2020irradiated}) rather than a state with a habitable liquid ocean. Of the 11
Hycean candidates in Table 1 of \cite{madhusudhan2021habitability}, all but one (K2-18b) is
above the runaway greenhouse threshold. For planets not too far above the threshold,
though, there is the possibility that cloud effects could suppress the runaway and permit
a liquid ocean.

\begin{figure}[ht!]
\plotone{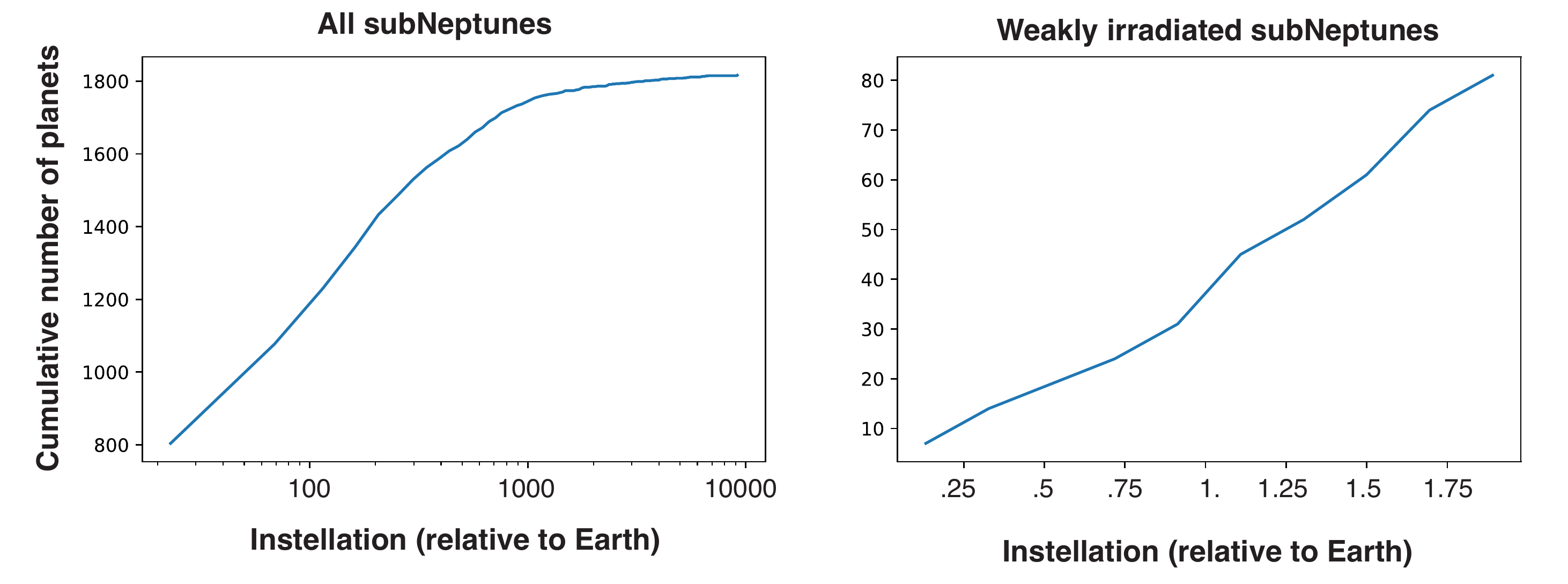}
\caption{Cumulative histograms of the instellation of all subNeptunes in the NASA Exoplanet Archive (left) and 
for the subset with instellation below twice that of the present Earth.    \label{fig:histograms}}
\end{figure}

It is of interest to see what candidates are available for investigation of systematic
differences between sub and super runaway subNeptune states. The precise instellation
threshold for runaway depends on stellar spectral class (which affects albedo), the
planet's surface gravity (a weak dependence), and cloud effects (a potentially profound
dependence), but as a general guideline, the threshold is in the vicinity of Earth's
present instellation. Fig.~\ref{fig:histograms} shows cumulative histograms of the
instellations of subNeptunes in the NASA exoplanet archive; subNeptunes are defined for
this figure as planets with an estimated radius between 1.5 and 3$r_\oplus$, and
instellations were computed from stellar effective temperature, stellar radius, and orbit
semi-major axis. Of the roughly 1800 subNeptunes in the archive, the overwhelming majority
are highly irradiated objects with intellation in excess of 30 times Earth's. These would
all be clearly in the super-runaway range, and most would be in regimes like that of
GJ1214b where water could not condense at any point in the atmosphere. There are, however,
80 confirmed planets with instellations below twice that of Earth, and this population
does straddle the runaway threshold. Only two of these (TOI-2257 b and TOI-2285 b) are
TESS detections, but it is to be hoped that TESS will add more candidates to this
population in the future. At present, only six of the 80 have measured radii and masses,
and three of those only have a weak upper bound on mass
(Table~\ref{table:MassRadPlanets}). The case of K2-18b has already been discussed. K2-3d
is super-runaway, but its radius is far smaller than what would be expected of a
super-runaway waterworld, so it is most likely a primarily rocky super-Earth, though at
80\% of Earth's mean density it most likely has a fairly substantial volatile envelope,
which could involve either supercritical water or hydrogen;future atmospheric characterization
could help discriminate between the two cases.  LHS1140b is sub-runaway but has radius much
smaller than the sub-runaway pure-water case; it is ultra-dense and most likely a
rock-ball with at most a thin atmosphere or ice/ocean shell. Kepler-138 b has a radius
in the range that is generally considered to be likely to be a rocky Super-Earth, but its
mass estimate of a mere .64$M_\oplus$ firmly positions it as a subNeptune, so we have included it in the table.
Its instellation puts it in the super-runaway category, unless the planet has an unusually high albedo. 
We did not include such low masses in Table \ref{table:massrad} because the low surface gravity of such
planets in the super-runaway state compromises our use of a fixed 280$\mathrm{W/m^2}$ runaway threshold,
but the estimated radius for a marginally super-runaway state using the nominal threshold is 2.3$r_\oplus$.
The large radius arises because of the weak gravity associated with low mass.  In reality this is an underestimate
of the radius,since with an estimated surface gravity of just 1.2$\mathrm{m/s^2}$ the runaway threshold 
would be considerably below 280$\mathrm{W/m^2}$, so the planet would be hotter and even more inflated.
Even the nominal estimate is nearly double the observed radius, so Kepler-138 b cannot be a super-runaway
waterworld with a supercritical interior.  If the albedo were large enough (and sufficiently
dominant over cloud greenhouse effects)  to put the planet in a sub-runaway state, then the estimated
radius falls to 1.25$r_\oplus$, which is similar to the observed value. However, Kepler-138 is a cool
M star, so it would be hard to achieve a high albedo given the weak Rayleigh scattering and low ice albedo
associated with M star illumination. It is more plausible that the planet consists of a small rocky core
surrounded by a deep $\mathrm{H_2}$ dominated envelope, but future atmospheric characterization of this
interesting target would help to resolve the matter. 

This does not leave us with any good candidates for runaway vs. sub-runaway water worlds,
though the possibility that K2-3d is a silicate/iron world with an extensive (though not
dominant) runaway atmosphere bears further investigation. Hopefully future discoveries and
mass measurements will add more candidates that could be fodder for analysis of the
runaway transition, but there is a long way to go
\begin{table}
\begin{tabular}{l|ccccccc}
             &K2-18b & Kepler-138 b &  K2-3d & Kepler-22b & Kepler-62e &LHS1140b & TOI-2285b\\
\toprule
Instellation & 1.24  &2.32          & 1.61   & 1.10       & 1.15       & .35     & 1.54 \\ 
$M$ &          8.92  &0.64          & 2.80   & $<$36      &$<$36       & 6.37    & $<$19.50\\
$r$ &          2.37  &1.21          & 1.52   & 2.38       & 1.61       & 1.64    & 1.73 \\
\end{tabular}
\caption{Instellation, mass and radii of weakly irradiated subNeptunes with observationally based mass estimates. All quantities
are given as multiples of Earth values.}
\label{table:MassRadPlanets}
\end{table}

Even though the observational implications of the runaway transition for interpretation of mass/radius data are
clouded by degeneracies and (at present) by lack of data,  the transition in interior state affects numerous
key planetary processes, such as generation of magnetic fields in conducting fluids or ices, and formation
of interior magma oceans.  Hence, the runaway transition indicates a significant transition in the behavior
of the population of subNeptunes, which will be increasingly important as more low-instellation subNeptunes
are discovered and characterized.

\begin{acknowledgments}
{\it Acknowledgment:}  This work was supported bythe European Research Council Advanced grant EXOCONDENSE, \#740963
\end{acknowledgments}

\facility{Exoplanet Archive}

%\bibliography{subNeptuneRunawayV2}{}

\begin{thebibliography}{}
\expandafter\ifx\csname natexlab\endcsname\relax\def\natexlab#1{#1}\fi
\providecommand{\url}[1]{\href{#1}{#1}}
\providecommand{\dodoi}[1]{doi:~\href{http://doi.org/#1}{\nolinkurl{#1}}}
\providecommand{\doeprint}[1]{\href{http://ascl.net/#1}{\nolinkurl{http://ascl.net/#1}}}
\providecommand{\doarXiv}[1]{\href{https://arxiv.org/abs/#1}{\nolinkurl{https://arxiv.org/abs/#1}}}

\bibitem[{Aguichine {et~al.}(2021)Aguichine, Mousis, Deleuil, \&
  Marcq}]{aguichine2021mass}
Aguichine, A., Mousis, O., Deleuil, M., \& Marcq, E. 2021, The Astrophysical
  Journal, 914, 84

\bibitem[{Bean {et~al.}(2021)Bean, Raymond, \& Owen}]{bean2021nature}
Bean, J.~L., Raymond, S.~N., \& Owen, J.~E. 2021, Journal of Geophysical
  Research: Planets, 126, e2020JE006639

\bibitem[{Benneke {et~al.}(2019)Benneke, Wong, Piaulet, Knutson, Lothringer,
  Morley, Crossfield, Gao, Greene, Dressing, {et~al.}}]{benneke2019water}
Benneke, B., Wong, I., Piaulet, C., {et~al.} 2019, The Astrophysical Journal
  Letters, 887, L14

\bibitem[{Blain {et~al.}(2021)Blain, Charnay, \& B{\'e}zard}]{blain20211d}
Blain, D., Charnay, B., \& B{\'e}zard, B. 2021, Astronomy \& Astrophysics, 646,
  A15

\bibitem[{Boukrouche {et~al.}(2021)Boukrouche, Lichtenberg, \&
  Pierrehumbert}]{boukrouche2021beyond}
Boukrouche, R., Lichtenberg, T., \& Pierrehumbert, R.~T. 2021, The
  Astrophysical Journal, 919, 130

\bibitem[{Budisa \& Schulze-Makuch(2014)}]{budisa2014supercritical}
Budisa, N., \& Schulze-Makuch, D. 2014, Life, 4, 331

\bibitem[{Charnay {et~al.}(2021)Charnay, Blain, B{\'e}zard, Leconte, Turbet, \&
  Falco}]{charnay2021formation}
Charnay, B., Blain, D., B{\'e}zard, B., {et~al.} 2021, Astronomy \&
  Astrophysics, 646, A171

\bibitem[{Dorn \& Lichtenberg(2021)}]{dorn2021hidden}
Dorn, C., \& Lichtenberg, T. 2021, The Astrophysical Journal Letters, 922, L4

\bibitem[{Guillot(2010)}]{guillot2010radiative}
Guillot, T. 2010, Astronomy \& Astrophysics, 520, A27

\bibitem[{Haldemann {et~al.}(2020)Haldemann, Alibert, Mordasini, \&
  Benz}]{haldemann2020aqua}
Haldemann, J., Alibert, Y., Mordasini, C., \& Benz, W. 2020, Astronomy \&
  Astrophysics, 643, A105

\bibitem[{Ingersoll(1969)}]{ingersoll1969runaway}
Ingersoll, A.~P. 1969, Journal of Atmospheric Sciences, 26, 1191

\bibitem[{Joshi \& Haberle(2012)}]{joshi2012suppression}
Joshi, M.~M., \& Haberle, R.~M. 2012, Astrobiology, 12, 3

\bibitem[{Kasting(1988)}]{kasting1988runaway}
Kasting, J.~F. 1988, Icarus, 74, 472

\bibitem[{Koll \& Cronin(2019)}]{koll2019hot}
Koll, D.~D., \& Cronin, T.~W. 2019, The Astrophysical Journal, 881, 120

\bibitem[{Li \& Pierrehumbert(2011)}]{li2011sea}
Li, D., \& Pierrehumbert, R.~T. 2011, Geophysical Research Letters, 38

\bibitem[{Madhusudhan {et~al.}(2021)Madhusudhan, Piette, \&
  Constantinou}]{madhusudhan2021habitability}
Madhusudhan, N., Piette, A.~A., \& Constantinou, S. 2021, The Astrophysical
  Journal, 918, 1

\bibitem[{Miller-Ricci \& Fortney(2010)}]{miller2010nature}
Miller-Ricci, E., \& Fortney, J.~J. 2010, The Astrophysical Journal Letters,
  716, L74

\bibitem[{Millot {et~al.}(2019)Millot, Coppari, Rygg, Correa~Barrios, Hamel,
  Swift, \& Eggert}]{millot2019nanosecond}
Millot, M., Coppari, F., Rygg, J.~R., {et~al.} 2019, Nature, 569, 251

\bibitem[{Misener \& Schlichting(2022)}]{misener2022importance}
Misener, W., \& Schlichting, H.~E. 2022, Monthly Notices of the Royal
  Astronomical Society, 514, 6025, \dodoi{10.1093/mnras/stac1732}

\bibitem[{Mol~Lous {et~al.}(2022)Mol~Lous, Helled, \&
  Mordasini}]{mol2022potential}
Mol~Lous, M., Helled, R., \& Mordasini, C. 2022, Nature astronomy, 6, 819

\bibitem[{Mousis {et~al.}(2020)Mousis, Deleuil, Aguichine, Marcq, Naar,
  Aguirre, Brugger, \& Gon{\c{c}}alves}]{mousis2020irradiated}
Mousis, O., Deleuil, M., Aguichine, A., {et~al.} 2020, The Astrophysical
  journal letters, 896, L22

\bibitem[{Nakajima {et~al.}(1992)Nakajima, Hayashi, \& Abe}]{nakajima1992study}
Nakajima, S., Hayashi, Y.-Y., \& Abe, Y. 1992, Journal of Atmospheric Sciences,
  49, 2256

\bibitem[{Nixon \& Madhusudhan(2021)}]{nixon2021deep}
Nixon, M.~C., \& Madhusudhan, N. 2021, Monthly Notices of the Royal
  Astronomical Society, 505, 3414

\bibitem[{Pierrehumbert \& Gaidos(2011)}]{pierrehumbert2011hydrogen}
Pierrehumbert, R., \& Gaidos, E. 2011, The Astrophysical Journal Letters, 734,
  L13

\bibitem[{Pierrehumbert(2010)}]{pierrehumbert2010principles}
Pierrehumbert, R.~T. 2010, Principles of planetary climate (Cambridge
  University Press)

\bibitem[{Pruteanu {et~al.}(2017)Pruteanu, Ackland, Poon, \&
  Loveday}]{pruteanu2017immiscible}
Pruteanu, C.~G., Ackland, G.~J., Poon, W.~C., \& Loveday, J.~S. 2017, Science
  Advances, 3, e1700240

\bibitem[{Scheucher {et~al.}(2020)Scheucher, Wunderlich, Grenfell, Godolt,
  Schreier, Kappel, Haus, Herbst, \& Rauer}]{scheucher2020consistently}
Scheucher, M., Wunderlich, F., Grenfell, J.~L., {et~al.} 2020, The
  Astrophysical Journal, 898, 44

\bibitem[{Shields {et~al.}(2013)Shields, Meadows, Bitz, Pierrehumbert, Joshi,
  \& Robinson}]{shields2013effect}
Shields, A.~L., Meadows, V.~S., Bitz, C.~M., {et~al.} 2013, Astrobiology, 13,
  715

\bibitem[{Tsiaras {et~al.}(2019)Tsiaras, Waldmann, Tinetti, Tennyson, \&
  Yurchenko}]{tsiaras2019water}
Tsiaras, A., Waldmann, I.~P., Tinetti, G., Tennyson, J., \& Yurchenko, S.~N.
  2019, Nature Astronomy, 3, 1086

\bibitem[{Turbet {et~al.}(2019)Turbet, Ehrenreich, Lovis, Bolmont, \&
  Fauchez}]{turbet2019runaway}
Turbet, M., Ehrenreich, D., Lovis, C., Bolmont, E., \& Fauchez, T. 2019,
  Astronomy \& Astrophysics, 628, A12

\bibitem[{von Paris {et~al.}(2013)von Paris, Selsis, Kitzmann, \&
  Rauer}]{von2013dependence}
von Paris, P., Selsis, F., Kitzmann, D., \& Rauer, H. 2013, Astrobiology, 13,
  899, \dodoi{10.1089/ast.2013.0993}

\bibitem[{Wordsworth \& Pierrehumbert(2013)}]{wordsworth2013water}
Wordsworth, R., \& Pierrehumbert, R. 2013, The Astrophysical Journal, 778, 154

\end{thebibliography}
%\bibliographystyle{aasjournal}



\end{document}